\documentclass[apjl]{emulateapj}

\newcommand{\kms}{{\rm km~s\ensuremath{^{-1}}}}

\newcommand{\beq}{\begin{equation}}
\newcommand{\eeq}{\end{equation}}

\newcommand{\lap}{\;\rlap{\lower 2.5pt \hbox{$\sim$}}\raise 1.5pt\hbox{$<$}\;}
\newcommand{\gap}{\;\rlap{\lower 2.5pt \hbox{$\sim$}}\raise 1.5pt\hbox{$>$}\;}

\def\mh{M_\bullet}

\usepackage{natbib}
\usepackage{multirow}
\usepackage{longtable}
\usepackage{amsmath}

\slugcomment{ApJL Accepted}

\shorttitle{A Displaced SMBH in M87}
\shortauthors{Batcheldor et al.}

\begin{document}


\title{A Displaced Supermassive Black Hole in M87
\footnote{B\lowercase{ased on observations made with the} NASA/ESA H\lowercase{ubble} S\lowercase{pace} 
T\lowercase{elescope, obtained at the} S\lowercase{pace} T\lowercase{elescope} S\lowercase{cience} 
I\lowercase{nstitute, which is operated by the} A\lowercase{ssociation of} U\lowercase{niversities for} 
R\lowercase{esearch in} A\lowercase{stronomy,} I\lowercase{nc., under} NASA \lowercase{contract} NAS5-26555.}}


\author{
D. Batcheldor\altaffilmark{1}, 
A. Robinson\altaffilmark{2}, 
D. J. Axon\altaffilmark{3,2},  
E. S. Perlman\altaffilmark{1}
\&
D. Merritt\altaffilmark{2}}
\email{dbatcheldor@fit.edu}


\altaffiltext{1}{Physics and Space Sciences Department, Florida Institute of Technology, 150 West University Boulevard, Melbourne, FL 32901, USA}
\altaffiltext{2}{Physics Department, Rochester Institute of Technology, 84 Lomb Memorial Drive, Rochester, NY 14623-5603, USA}
\altaffiltext{3}{School of Mathematical and Physical Sciences, University of Sussex, Sussex House, Brighton, BN1 9RH, UK}

\begin{abstract}
Isophotal analysis of M87, using data from the Advanced Camera for Surveys, reveals a
projected displacement of $6.8\pm0.8$~pc ($\sim0\farcs1$) between the nuclear point source 
(presumed to be the location of the supermassive black hole, SMBH) and the photo-center of
the galaxy. The displacement is along a position angle of $307\pm17$\degr\ and is consistent 
with the jet axis. This suggests the active SMBH in M87 does not currently reside at the galaxy 
center of mass, but is displaced in the counter-jet direction. Possible explanations for the displacement 
include orbital motion of an SMBH binary, gravitational perturbations due to massive objects (e.g., 
globular clusters), acceleration by an asymmetric or intrinsically one-sided jet, and gravitational 
recoil resulting from the coalescence of an SMBH binary. The displacement direction favors the latter 
two mechanisms. However, jet asymmetry is only viable, at the observed accretion rate, for a jet age of 
$>$0.1~Gyr and if the galaxy restoring force is negligible. This could be the case in the low density 
core of M87. A moderate recoil $\sim$1~Myr ago might explain the disturbed nature of the nuclear 
gas disk, could be aligned with the jet axis, and can produce the observed offset. Alternatively, the
displacement could be due to residual oscillations resulting from a large recoil that occurred in the
aftermath of a major merger $\le$1~Gyr ago.
\end{abstract}


\section{Introduction}

It is generally assumed that supermassive black holes (SMBHs) reside at the centers of their host galaxies. However, SMBHs 
can be significantly displaced from their central locations by asymmetric forces during a merger, or by a second SMBH \citep{2006MmSAI..77..733K}. 
In addition, if a binary SMBH forms and coalesces, anisotropic emission of gravitational waves can result in initial impulsive kick velocities 
of several thousand \kms \citep[e.g.,][]{2007arXiv0710.1338P}. Even if the kick velocity is small enough for the 
coalesced SMBH to remain in the galaxy, {\it N}-body simulations have shown that the SMBH can oscillate within the bulge for $\sim1$~Gyr before coming 
to rest \citep{2008ApJ...678..780G}. Alternatively, if the galaxy contains a radio source, the SMBH may experience sustained acceleration due to intrinsic 
asymmetries in jet power \citep[e.g.,][]{2007ARep...51...97T}. 

The most direct way to find a displaced SMBH is to observe a spatial offset between the SMBH and the center of its host galaxy. This requires data at 
the highest possible spatial resolution. E/S0 galaxies that are minimally affected by extinction and contain an active galactic nucleus (AGN) provide 
good candidates; bulge isophotes determine the position of the galaxy center and the AGN (point source) determines the position of the SMBH. M87 is 
an ideal target for a displaced SMBH search. It is nearby, has a regular bulge, is relatively free of dust, hosts an AGN and jet \citep[e.g.,][]{2001ApJ...551..206P}, 
and has been extensively observed by the Hubble Space Telescope ({\it HST}). In this {\it Letter}, we report the discovery of a $6.8\pm0.8$~pc projected 
displacement between the center of M87, as defined by the galaxy isophotes, and 
the SMBH. 

\section{The Data}

Our analysis uses the archived {\it HST} data listed in Table~\ref{tab:1}. All data had the standard STScI on-the-fly re-processing 
applied. Each image was rotated to the same reference frame and shifted to a common position using a 2D cross correlation register. The data were 
combined using a median filter to remove residual cosmic rays and bad pixels, and to minimize the effect of the High Resolution Channel (HRC) 
coronagraphic aberration. 

\begin{figure}
\plotone{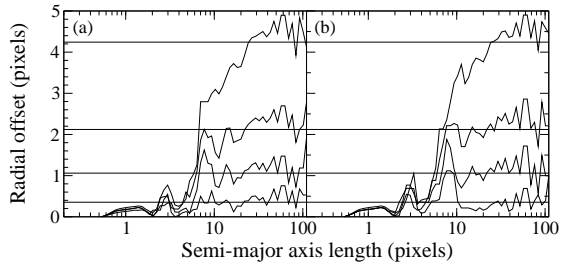}
\caption{Accurate recovery of simulated SMBH-galaxy displacements. {\it [a]} Without {\it [b]} with a mask. Horizontal lines show the model 
displacements. Offsets are recovered for SMA$<$10 pixels. 
}
\label{fig:1}
\end{figure}

The IRAF task ELLIPSE \citep{1987MNRAS.226..747J} was used to determine the photo-center of the galaxy. Beginning with a semi-major axis (SMA) of 1~pixel, 
centered on the nuclear point source, ellipses of progressively increasing SMA were independently fitted to the data. The SMA was incremented by 1~pixel in 
each successive fit and the center of the ellipse found. The $x-y$ pixel co-ordinates of the ellipse centers were thus determined as a function of SMA.
To estimate the precision with which offsets can be recovered, and check whether masking degrades accuracy, we applied this technique 
to a set of simulated galaxies containing a nuclear point source. The simulated galaxies were given an $r^{1/4}$ surface brightness profile and an ellipticity of 
0.2. Point sources, with offsets of 0, 0.25, 0.5, 0.75, 1.0, 1.5, 2.0, 3.0 pixels in both $x$ and $y$, were added to create 64 total models. Each model 
was convolved with the {\it HST} point spread function (PSF) as generated by TinyTim \citep{1995ASPC...77..349K}, and populated with random noise. First, 
isophotal fits were performed without a mask. In this case, the offsets were recovered to $\sim$0.2 pixels 
(Fig.~\ref{fig:1}a). Second, isophotes were fitted with a mask that simulated an extended jet crossing the galaxy center, multiple globular clusters 
and areas of extinction. These offsets were also recovered to $\sim$0.2 pixels (Fig.~\ref{fig:1}b); any observed offsets will not be a result 
of masking. 

\begin{figure}
\plotone{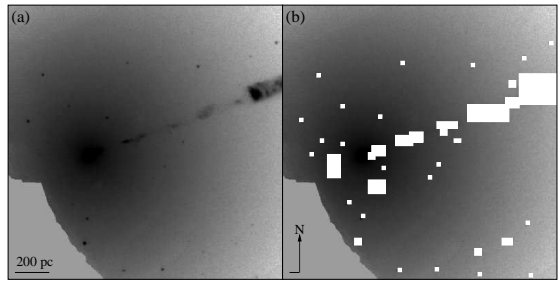}
\caption{{\it [a]} ACS HRC F814W median combined image of M87 with a logarithmic stretch. {\it [b]} The mask overlaid on {\it [a]}.
}
\label{fig:2}
\end{figure}

Figure~\ref{fig:2} shows the median combined HRC data in F814W, both with and without the mask. A distance modulus of $31.0\pm0.2$ 
\citep{2001ApJ...546..681T} puts M87 at $16.1\pm0.2$ Mpc (77.9~pc per arcsec). The ACS pixel scales of 0\farcs027 (HRC) and 
0\farcs049 (Wide Field Channel, WFC) are then 2.1 and 3.8~pc, respectively. 

\begin{deluxetable}{lcccc}
\tablecaption{Archived ACS Data \label{tab:1}}
\tablewidth{0pt}
\tablehead{
\colhead{Dataset}&
\colhead{Aperture}&
\colhead{Filter}&
\colhead{Exposure times}&
\colhead{PID}
}
\startdata
j8q0* & HRC & F606W & $2\times75$s              & 9829  \\
j92j* & HRC & F606W & $4\times80$s,$5\times48$s & 10133 \\
j9ei* & HRC & F606W & $6\times80$s,$3\times45$s & 10617 \\
j9qf* & HRC & F606W & $2\times80$s,1$\times45$s & 10910 \\
j8l0* & HRC & F814W & $5\times100$s 		& 9705  \\
j8q0* & HRC & F814W & $8\times50$s   		& 9829  \\
j92j* & HRC & F814W & $4\times50$s   		& 10133 \\
j9ei* & HRC & F814W & $6\times48$s   		& 10617 \\
j9qf* & HRC & F814W & $2\times48$s   		& 10910 \\
j9e0* & WFC & F606W & $3\times500$s 		& 10543 \\
j9e0* & WFC & F814W & $3\times1440$s 		& 10543 \\
\enddata
\tablecomments{Details of the archived ACS data used. PID is the {\it HST} program ID number.
}
\end{deluxetable}

\section{Results}

\begin{figure*}
\plotone{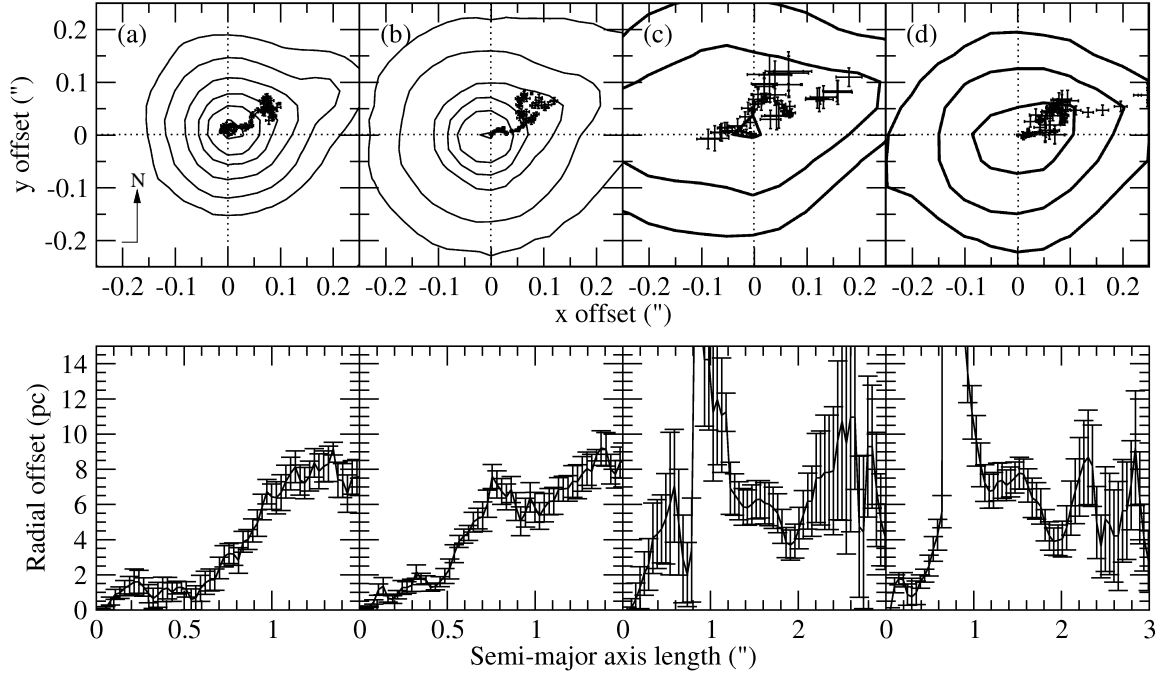}
\caption{{\it[Top row]} $x$ and $y$ ellipse centers overlaid on ACS contours. The point source is at [0,0]. {\it [a,b]} HRC ellipse centers for SMA$<$1\farcs5. 
The jet (masked out for the fits) can be seen as a N-W extension. {\it [c,d]} WFC ellipse centers for SMA$<$3\farcs0. {\it [a,c]} F606W 
{\it [b,d]} F814W. {\it [Bottom row]} Radial offsets of the ellipse centers.}
\label{fig:3}
\end{figure*}

Table~\ref{tab:2} presents the ellipse fits to the HRC and WFC data in F606W and F814W. The ellipticity and position angle (PA) of the fits are 
uncorrelated with SMA, show no evidence for bulge asymmetry, and are consistent with the results of \cite{2006ApJS..164..334F} who note the isophotes 
are regular.
At each SMA the radial offset of 
the ellipse center is given with respect to the nuclear point source. Figure~\ref{fig:3} presents the $x$ and $y$ co-ordinates of the ellipse centers 
and the radial offset as a function of SMA. Note that the inner jet region is seen 
as ``pear shaped'' contours in Figure~\ref{fig:3}(a,b). This feature was masked out during the fits. The elongated contours in  Figure~\ref{fig:3}(c,d) 
are an effect of charge bleeding from the strong point source, and result in larger uncertainties and the anonymously large offsets seen at $\sim$1\farcs0.

All ellipses with SMA$>$1\farcs0 show a clear offset relative to the nuclear point source. For SMA$<$1\farcs0, the measured offsets show the transition between 
isophotes dominated by the AGN and the bulge.

The mean radial offset from both the HRC and WFC (1\farcs0$<$SMA$<$3\farcs0), weighted by the uncertainties, is $6.84\pm0.07$~pc. 
The error on the mean is substantially less than the 0.2 pixel uncertainty implied by the simulations (corresponding to 0.76~pc in the WFC data), which we prefer 
to adopt as a conservative estimate of the offset uncertainty.  
Therefore, the best estimate of the radial offset is $6.8\pm0.8$~pc. The uncertainty weighted PA between the offset and the point source is 
$306\pm17\degr$, consistent with the jet axis \citep[e.g.][]{1980ApJ...239L..11O}. 

\section{Origin of the Displacement}

The observed offset between the bulge photo-center and the AGN suggests that the SMBH in M87 is displaced from the stellar center of mass. The 
projected displacement is $\sim 7$~pc ($\sim0\farcs1$) approximately in the {\em counter-jet} direction. Here we consider several mechanisms 
that might produce this displacement.

\begin{deluxetable}{ccc|ccc}
\tablecaption{Radial Offsets of Ellipse Centers\label{tab:2}}
\tablewidth{0pt}
\tablehead{
\multicolumn{3}{c}{HRC}&
\multicolumn{3}{c}{WFC}\\ \\
\colhead{SMA ($\arcsec$)}&
\multicolumn{2}{c}{Offset (pc)}&
\colhead{SMA ($\arcsec$)}&
\multicolumn{2}{c}{Offset (pc)}\\ \\
\colhead{}&
\colhead{F606W}&
\colhead{F814W}&
\colhead{}&
\colhead{F606W}&
\colhead{F814W}
}
\startdata
0.027 & 0.000 & 0.000 & 0.049 & 0.000 & 0.000 \\
0.054 & 0.110 & 0.187 & 0.098 & 0.122 & 1.405 \\
0.081 & 0.513 & 0.453 & 0.147 & 0.602 & 1.779 \\
0.108 & 0.889 & 1.155 & 0.196 & 1.272 & 1.718 \\
0.135 & 0.982 & 1.350 & 0.245 & 2.071 & 0.870 \\
\enddata
\tablecomments{SMA verses radial offset for all ACS data. Table~\ref{tab:2} is published in its entirety in the electronic edition of 
the ApJ Letters. 
}
\end{deluxetable}

\subsection{Jet Asymmetry} 

The PA of the displacement suggests a connection with the M87 jet.
\cite{1982IAUS...97..475S} first noted that one-sided jets can accelerate SMBHs. However, the standard interpretation for one-sided jets 
is relativistic beaming \citep[e.g.,][]{1983Natur.303..779E}. In the case 
of M87, there is VLBI and VLBA evidence for a probable counter-jet \citep{2007ApJ...660..200L,2007ApJ...668L..27K}, the large-scale radio structure 
shows evidence for two-sided jet activity in earlier epochs \citep{2000ApJ...543..611O}, and Chandra images show symmetric X-ray inner cocoons 
\citep{2007ApJ...665.1057F} that could be used to constrain the jet asymmetry. 
Models in which the SMBH is accelerated by two-sided, but intrinsically asymmetric, jets have been explored by \cite{1992ApJ...390...46W} 
and \cite{2007ARep...51...97T}.
In this case, the SMBH acceleration is 
\beq
a_\mathrm{BH} \approx 2.1\times 10^{-6} f_\mathrm{jet}\dot m\ 
\mathrm{cm}\ \mathrm{s}^{-2}
\eeq
\citep{2008ApJ...681..104K} where $f_\mathrm{jet}$ is the luminosity of the asymmetric part of the jet, expressed as a fraction 
of the accretion luminosity. The mass accretion rate ($\dot{m}$) is in units of $\dot{M}_\mathrm{Edd}=L_\mathrm{Edd}/(\epsilon c^2)$, 
where $\epsilon$ is the accretion efficiency.

As M87 has a large, low-density core \citep[e.g.,][]{1992AJ....103..703L,2006ApJS..164..334F}, we first assume the restoring force 
from the galaxy is negligible.
The displacement ($\Delta{r}$) and velocity ($\Delta{v}$) 
of the SMBH would increase with time as
\begin{subequations}
\begin{eqnarray}
\Delta r &\approx& 340 \mathrm{pc}\, f_\mathrm{jet}\dot m\, t_6^2, \\
\Delta v &\approx& 660 \mathrm{km\ s}^{-1} f_\mathrm{jet}\dot m\, t_6
\end{eqnarray}
\label{eq:norestore}
\end{subequations}
\noindent where $t_6$ is the time in Myr since the jet turned on. 
Assuming the SMBH is offset in the counter-jet direction, a projected displacement of 
$6.8\pm0.8$~pc, and a jet orientation of $15\pm5\degr$ \citep{1999ApJ...520..621B}, gives a physical $\Delta{r}$ of $26^{+18}_{-9}$~pc. However, 
this could be as small as $\sim10$~pc for a jet orientation of $45\degr$ \citep{2007ApJ...660..200L}. An upper limit to the jet lifetime is set by age 
of the outer radio halo \citep[$\sim~0.1$Gyr;][]{2000ApJ...543..611O}, assuming it is still powered by the jet. Alternatively, if the outer halo is a 
relic, then a lower limit to the jet lifetime is given by the age of the inner lobes \citep[$\sim$1~Myr;][]{1996ApJ...467..597B}. 
Equations~(\ref{eq:norestore}) therefore give
\begin{subequations}
\begin{eqnarray}
&3\times 10^{-6}\left(\frac{\Delta r}{10\mathrm{pc}}\right) \le f_\mathrm{jet}\dot m \le 3\times 10^{-2}\left( \frac{\Delta r}{10\mathrm{pc}}\right),\\
&0.2\left(\frac{\Delta r}{10\mathrm{pc}}\right) \kms \le \Delta{v} \le 20 \left(\frac{\Delta r}{10\mathrm{pc}}\right)\kms .
\end{eqnarray}
\label{eq:norestore2}
\end{subequations}

\cite{2003ApJ...582..133D} report $\dot{m}\approx~10^{-4}$ in M87. A 1~Myr jet then requires $f_\mathrm{jet}~>>~1$ and can be ruled out. 
However, for a $\sim$0.1~Gyr jet, $f_{jet}\approx 0.03$, i.e., the jet asymmetry amounts to only $\sim 3$\% of the 
accretion luminosity. Therefore, the displacement could result from acceleration by a long-lived jet 
if the galaxy restoring force is negligible. 

In the presence of a restoring force, the SMBH is expected to come to rest where the force from the galaxy matches the jet force. Adopting units such that 
$G=r_c=\sigma=1$, where $r_c\approx{500}$~pc is the M87 core radius and $\sigma\approx 330\kms$ is the 1d core stellar velocity dispersion, the 
equation of motion of an accelerated SMBH in a fixed galaxy core is \citep[e.g.][]{2008ApJ...678..780G}
\begin{equation}\label{eq:motion}
\ddot{x_i} + 2\gamma\dot{x_i} + \beta^2x_i = a_i 
\end{equation}
where the $x_i$ are Cartesian coordinates and
\begin{subequations}
\begin{eqnarray}
\gamma &\equiv& \frac{9}{\sqrt{2\pi}} F \frac{\mh\ln\Lambda}{M_c}\\
\beta^2 &\equiv& 3C_i \equiv\omega_i^2 \\
a_i &\equiv& - 3.6 \frac{r_{500}}{\sigma_{300}^2} f_\mathrm{jet}\dot m 
\lambda_i.
\end{eqnarray}
\end{subequations}
\noindent
The third term on the LHS of equation~\ref{eq:motion} represents the restoring force on the displaced SMBH from the stars, assuming that the motion takes 
place in a homogeneous core; the $C_i$ are related to the shape of the core and are unity for a spherical galaxy \citep[e.g.,][]{1987QB410.C47......}. The 
second term represents the dynamical friction force from the stars and has been expressed in terms of the core mass, $M_c\equiv{4}\pi\rho{r_c}^3/3$; $\ln\Lambda$ 
is the Coulomb logarithm and $F\lap{1}$ is a ``fudge factor'' accounting for the fact that the frictional force on massive objects in galaxy cores is found 
to be somewhat less than predicted by Chandrasekhar's formula \citep{2008ApJ...678..780G,2009arXiv0912.2409I}. Finally, $\lambda_i=\hat{e_j}\cdot\hat{e_i}$ 
where $\hat{e_j}$ is a unit vector in the direction of the jet, and $r_{500}\equiv{r_c}/(500\mathrm{pc})$, $\sigma_{300}\equiv\sigma/(300\mathrm{\kms})$.

In the absence of dynamical friction ($\gamma=0$), the solutions to equation~(\ref{eq:motion}) are
\beq
x_i(t) = -X_i\left(\cos\omega_it - 1\right),\ \ \ \ 
X_i = \frac{a_i}{\omega_i^2},
\eeq
\noindent
i.e. continued oscillation about the point $\mathbf{X}$ where the jet acceleration is balanced by the gravitational force. Unless the core is very non-spherical, 
$\mathbf{X}$ will point approximately opposite the jet direction.

In the presence of dynamical friction, the oscillations are damped and asymptote to $\mathbf{X}$. Furthermore, since $M_\bullet$ is of order $M_c$, 
the damping time is expected to be comparable to the core crossing time, $T_c=r_c/\sigma_c\approx1.5~\mathrm{Myr}~(r_{500}/\sigma_{300})$. 
This is somewhat larger if $F<1$, but is nevertheless comparable with the lower limit on the jet lifetime set by the age of the inner lobes.
 
Reintroducing dimensional variables,

\begin{equation}\label{eq:displace}
\Delta r\simeq 500 \mathrm{pc} \left(f_\mathrm{jet}\dot m\right)r_{500}^2\sigma_{300}^{-2}.
\end{equation}
\noindent

This implies $0.02 \lap f_\mathrm{jet}\dot m \lap 0.05$ for $10 \mathrm{pc} \lap\Delta r\lap 25 \mathrm{pc}$. 
Even assuming $f_\mathrm{jet}=1$ (one-sided jet), the observed offset requires $\dot m\approx 0.02$, i.e., several magnitudes more than 
observed. In addition, while the AGN in M87 may have been more active in the past, the offset requires the jet 
to have remained at this high level of activity for $\gtrsim$1~Myr.

\subsection{Binary SMBH} 

If the SMBH in M87 is one component of a bound pair, with mass ratio $q\equiv M_2/M_1 \le 1$, then
\begin{equation}
\Delta r \approx q \Delta r_2 \approx 10 \mathrm{pc} \left(\frac{q}{0.1}\right)\left(\frac{\Delta r_2}{100 \mathrm{pc}}\right)
\end{equation}
where $\Delta r_2$ is the separation of the smaller SMBH from the binary center of mass (located at the galaxy center). A binary separation of $\sim 100$~pc 
is not unreasonable given the high merger rate expected for a luminous galaxy at the center of a rich cluster, and given the gradual rate of in-spiral expected 
in the low-density core. 

The velocity of the 
larger SMBH with respect to the binary center of mass (assuming a circular orbit) would be much greater than in the jet scenario:
\begin{subequations}\label{model2}
\begin{eqnarray}
&V_1=\frac{q}{1+q}\sqrt{\frac{G(M_1+M_2)}{a}} \\
&\approx 400\kms \frac{q}{1+q}\left(\frac{M_1+M_2}{4\times 10^9M_\odot}\right)^{1/2}\left(\frac{a}{110 \mathrm{pc}}\right)^{-1/2}
\end{eqnarray}
\end{subequations}
\noindent
where $a$ is the binary separation. However, the line-of-sight velocity is $V_\mathrm{los} = V_1\sin i\sin \phi$, where $i$ and $\phi$ are the 
unknown inclination and phase of the orbit. If the nuclear gas disk is in the orbital plane then $i\approx 50\degr$ \citep{1997ApJ...489..579M} 
which gives $V_\mathrm{los} \approx 220\kms$ when $\phi = 45\degr$. 

A non-active second SMBH would be extremely hard to detect, but a likely
consequence of such a binary 
would be jet precession \citep[e.g.,][]{2000A&A...360...57R}. Several jet knots have strong helical morphologies, but their widths 
are consistent with a steadily expanding cone with near-zero width at the nucleus \citep{2005ASPC..340..104L}. Since there is no evidence of a 0\farcs1 scale 
``wobble'' of the jet direction in both {\it HST} and VLBA data, any precession must be several orders of magnitude smaller than that expected from a binary.

\subsection{Massive Perturbers} 

Any SMBH will experience gravitational perturbations from stars and more massive objects, e.g. globular clusters (GCs), open clusters, etc. The result is a 
Brownian motion of the SMBH, with mean square velocity
\beq
\frac{1}{2}\mh\langle V_\mathrm{SMBH}^2\rangle \approx \frac{3}{2}\tilde m \tilde\sigma^2.
\eeq
\noindent
Here, $\tilde{m}$ is the second moment of the mass distribution of perturbing objects, and $\tilde\sigma$ is the velocity dispersion of the perturbers measured 
within $\sim0.5r_i$ ($r_i$ is the SMBH's influence radius, \citealt{2007AJ....133..553M}). The rms displacement of the SMBH is then given by the virial theorem:
\begin{equation}\label{eq:rrms}
\Delta r_\mathrm{rms} \approx 
\left(\frac{\tilde m}{\mh}\right)^{1/2} r_c.
\end{equation}
\noindent
Assuming that GCs constitute a fraction $\sim 10^{-3}$ by mass of M87 \citep{1994ApJ...422..486M} and that the mass of one GC is $10^5M_\odot$, 
Equation~\ref{eq:rrms} implies $r_\mathrm{rms}\lap 0.1$~pc. This is likely an overestimate since the density profile of GCs is flatter than 
that of the galaxy light. Unless there is another population of ``massive perturbers'' in M87 \citep{2007ApJ...656..709P} it is unlikely that 
the observed displacement can be due to gravitational perturbations.

\subsection{Gravitational Wave Recoil} 

A kick velocity of $v_\mathrm{k}\approx~500$\kms\ would displace the coalesced SMBH a distance $\sim r_c$ 
from the center of M87, in a direction orthogonal to the orbital plane of the preceding binary \citep{2010arXiv1003.3865V}, 
and aligned with the jet axis. 
Such a velocity is easily produced during the coalescence of 
two modestly spinning, comparably massive SMBHs \citep[e.g.,][]{2007PhRvD..76f1502T}. Larger kicks (but 
smaller than the escape velocity) would result in SMBH-core oscillations that damp on a time scale
\begin{equation}
T_\mathrm{damp}\approx 15 \frac{\sigma^3}{G^2\rho M_\bullet} 
\approx 2\times 10^8 \mathrm{yr}\ \sigma_{300}^{-3.86} r_{500}^2
\end{equation}
\citep{2008ApJ...678..780G}. The rms displacement of the SMBH with respect to the galaxy center decreases with time as
\begin{equation}
r_\mathrm{rms}(t)\approx r_c e^{-(t-t_c)/2T_\mathrm{damp}}, \ \ \ 
t>t_c
\end{equation}
where $t_c$ is the time at which the amplitude of the oscillations has damped to a scale of $\sim r_c$. A current displacement of $\sim 20$~pc 
$\approx 0.05r_c$ implies that $\sim 6T_\mathrm{damp}\approx 10$~Gyr have elapsed since $t_c$. Thus, a major SMBH merger during the formation 
of M87 could have generated a kick consistent with the current displacement. 

\section{Discussion}
Four displacement scenarios have been considered; jets, binaries, perturbers, and gravitational recoil. 
Perturbers cannot produce the observed displacement amplitude, and the observed jet-offset alignment 
must be fortuitous in the binary case. 
On the other hand, jet acceleration and gravitational 
recoil provide natural explanations for the jet-offset alignment.

These mechanisms predict displacement velocities that are either $\lap10\kms$ or $\gap100\kms$. For example, 
jet acceleration of the SMBH produces an rms velocity
$v_\mathrm{rms}~\approx~0\kms$ in the 
presence of a restoring force, and $\Delta{v}~\approx~20\kms~t_6^{-1}$ without. 
For SMBH-core oscillations induced by a large and early kick,
the rms velocity of the SMBH is
\begin{equation}
v_\mathrm{rms}\approx \sigma \frac{r_\mathrm{rms}}{r_c} 
\approx 12\ \mathrm{km\ s}^{-1} \sigma_{300} r_{500}^{-1} 
\frac{r_\mathrm{rms}}{20\ \mathrm{pc}},
\end{equation}
independent of the time since the kick or the kick amplitude. However, a velocity of several hundred \kms\ is possible if 
the SMBH is undergoing a large amplitude oscillation following a recent ($\le 1$ Myr) kick. A similar velocity is expected 
in the binary SMBH scenario. 

It may be possible to distinguish between the high and low velocity displacement 
mechanisms by comparing the relative recessional velocities of stars at the galaxy 
center and gas around the SMBH.
Unfortunately, this will be difficult; the low central surface brightness will limit 
the accuracy with which velocities can be determined from stellar absorption lines. Alternatively, 
wide field VLBA astrometry, referenced against background quasars 
\citep{2009AAS...21342009D}, may allow the proper motion of the nuclear point source to be determined. 
However, it will be difficult to distinguish between the cluster motion of M87 and the intrinsic kinematics of the SMBH. 

The central gas disk 
\citep{1994ApJ...435L..35H,1994ApJ...435L..27F}
may provide 
alternative constraints on the displacement velocity.
For example, a displaced SMBH retains gas whose Keplerian velocity exceeds the kick 
velocity \citep{2006MNRAS.367.1746M}. 
\cite{1994ApJ...435L..27F} estimate the disk-like structure in M87 to extend to $\sim80$~pc. 
Assuming this represents the portion of the disk that remains bound to the 
SMBH, we can infer an initial kick velocity $\sim 400\kms$. 
This would produce strongly shocked regions in the disk, as it passes 
through the ambient gas, that may explain several observed properties.
For example, while the gas in the central $\pm0\farcs4$ 
region of the disk is photo-ionized \citep{1997ApJ...489..579M,1999ApJ...527..733S} and has a closely Keplerian velocity field 
\citep{1994ApJ...435L..35H,1997ApJ...489..579M}, at larger radii non circular motions and morphological distortions are present 
\citep{1999LNP...530..278F,2004AAS...20511002B}. In addition, gas in the outer parts of the disk is evidently excited by a 265\kms shock 
\citep{1997ApJ...490..202D}, rather than the central UV continuum source. 

Two further consequences from a kick velocity $v_\mathrm{k}\approx400\kms$ are noted. 
First, the shocked gas may be hot enough ($T\sim~2\times~10^6$\,K) to
produce X-ray emission. Second, at this velocity the observed displacement is 
reached after $\sim~25000(\Delta{r}/10\mathrm{pc})$~yr, i.e., significantly less than 
the minimum jet age discussed earlier. If this minimum jet age corresponds to the time
since the kick, the SMBH cannot be on its initial outward trajectory. 
\cite{2008ApJ...678..780G} show for $v_k<50\%$ of the escape velocity, 
the SMBH motion is rapidly damped and undergoes only one large amplitude oscillation on a 
time-scale of 1~Myr . Therefore, it is possible the SMBH is on a return trajectory.

\section{Summary}

We find that the SMBH in M87 is displaced relative to the galaxy photo-center by
a projected distance of $6.8\pm0.8$~pc ($\sim0\farcs1$) in the
counter-jet direction. Four explanations have been considered: jet-induced acceleration, a
binary SMBH, massive perturbers and a gravitational wave kick. Perturbers only produce
offsets of $\sim0.1$~pc, and precession of an SMBH binary would 
produce ``wiggles'' in the jet that are
not seen. Neither can explain the observed alignment with the jet axis. 
Jet acceleration cannot be ruled out, but it requires both that the jet age to be $>>1$~Myr and 
that the restoring force exerted by the galaxy is small. The displacement could also be be explained 
by a moderate (a few hundred \kms) kick which occurred $\sim1$~Myr ago. This could produce the alignment 
with the jet axis, and may also explain the disturbed nature of the nuclear gas disk.
Alternatively, the displacement may be from SMBH-core oscillations following a kick that 
occurred $\le$1~Gyr ago.

The displacement processes discussed are likely to be common in early type galaxies; they are partly assembled via mergers and often host AGN-powered
radio sources. It is plausible, therefore, that there is a high incidence of SMBH displacements among the E/S0 population. A systematic effort to 
determine the statistical distribution of displaced nuclei will provide important insights into merger histories, the frequency with 
which SMBH binaries form and coalesce, and/or jet launch physics. Whatever the mechanism, displaced SMBHs have important consequences 
for accretion flows, for stellar dynamics in galactic nuclei, and for our fundamental understanding of galaxy evolution. 

\acknowledgments

We thank the referee for comments that improved the clarity of this paper. Support for this work was provided by proposal number HST-AR-11771.01 awarded by 
NASA through a grant from the Space Telescope Science Institute, which is operated by the Association of Universities for Research in Astronomy, Incorporated, 
under NASA contract NAS5-26555. DM was supported by grants AST-0807910 (NSF) and NNX07AH15G (NASA).

\end{document}